\documentclass[12pt]{article}
\usepackage{graphicx,subcaption}

\newcommand\pubnumber{arXiv}
\newcommand\pubdate{\today}

\def\institute{Physics Department and INFN,\\
Universit\`a di Milano Bicocca, 20146 Milano, ITALY}
\def\support{
}

\def\Title#1{\begin{center} {\Large #1 } \end{center}}
\def\Author#1{\begin{center}{ \sc #1} \end{center}}
\def\Address#1{\begin{center}{ \it #1} \end{center}}

\newcommand\pubblock{\rightline{\begin{tabular}{l} \pubnumber\\
         \pubdate  \end{tabular}}}
\newenvironment{Abstract}{\begin{quotation}  }{\end{quotation}}
\newenvironment{Presented}{\begin{quotation} \begin{center} 
             PRESENTED AT\end{center}\bigskip 
      \begin{center}\begin{large}}{\end{large}\end{center} \end{quotation}}

\def\beq{\begin{equation}}
\def\eeq#1{\label{#1}\end{equation}}
\def\eeqn{\end{equation}}

\def\beqa{\begin{eqnarray}}
\def\eeqa#1{\label{#1}\end{eqnarray}}
\def\eeqan{\end{eqnarray}}

\let\bar=\overbar

\def\etal{{\it et al.}}

\def\Dslash{\not{\hbox{\kern-4pt $D$}}}
\def\dslash{\not{\hbox{\kern-2pt $\del$}}}

\def\msb{{\bar{\ssstyle M \kern -1pt S}}}

\begin{document}
\begin{titlepage}
\pubblock

\vfill
\Title{Renormalons and the Top Quark Mass Measurement}
\vfill
\Author{Paolo Nason\support}
\Address{\institute}
\vfill
\begin{Abstract}
  I illustrate a recent work on the large-order behaviour of the
  perturbative expansion (and the related power-suppressed ambiguitiers)
  arising from infrared renormalons,
  in the context of top mass measurements in open-top production processes.
\end{Abstract}
\vfill
\begin{Presented}
$11^\mathrm{th}$ International Workshop on Top Quark Physics\\
Bad Neuenahr, Germany, September 16--21, 2018
\end{Presented}
\vfill
\end{titlepage}
\def\thefootnote{\fnsymbol{footnote}}
\setcounter{footnote}{0}

\section{Introduction}
A major worry in top mass measurements at Hadron Colliders has to do with
linear power corrections, i.e. non perturbative effects that are proportional
to a typical hadronic scale. In fact, the current experimental errors, reaching
values of the order of several hundreds MeV, are themselves of the order
of typical hadronic scales, and thus, our lack of a full understanding
of QCD at low energy is a source of concern regarding the value of
the measurement and of its associated error.
At the moment, the only methods at our disposal for estimating non-perturbative effects
is to vary the parameters and settings of the hadronization model,
and of all
the parameters that control the end of the Monte Carlo shower and the onset
of hadronization phenomena. Yet, since we are only dealing with models, the
doubt that we may not be covering the behaviour of the real physics
is hard to dismiss\footnote{For a recent discussion of these issues, see
 sec. 6.5.1 of ref.~\cite{CERN-LPCC-2018-03} and the contribution
  of A. Hoang to these proceedings.}.
It is therefore useful to consider simplified
theoretical frameworks where at least some aspects of the non-perturbative
corrections can be fully understood. One such framework is the large-$b_0$
approximation~\cite{Beneke:1994qe,
  Ball:1995ni},
where corrections corresponding to the insertion
of a light quark loop in the gluon propagator, as well as those due to a final
state gluon splitting into a quark-antiquark pair, are considered up to all orders in the
coupling constant.

\begin{figure}[htb]
 \centering
 \includegraphics[width=0.24\textwidth]{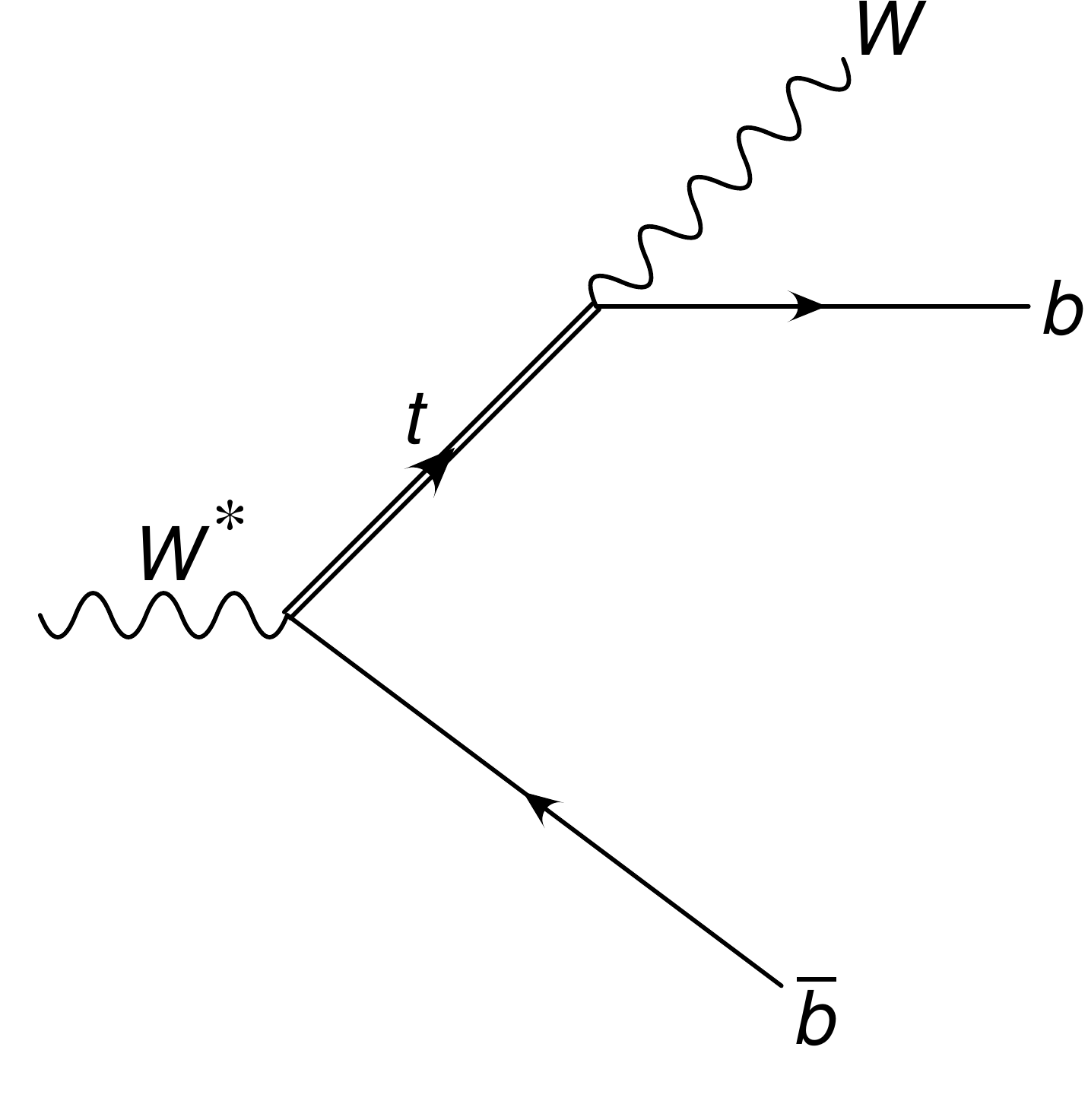}
 \includegraphics[width=0.24\textwidth]{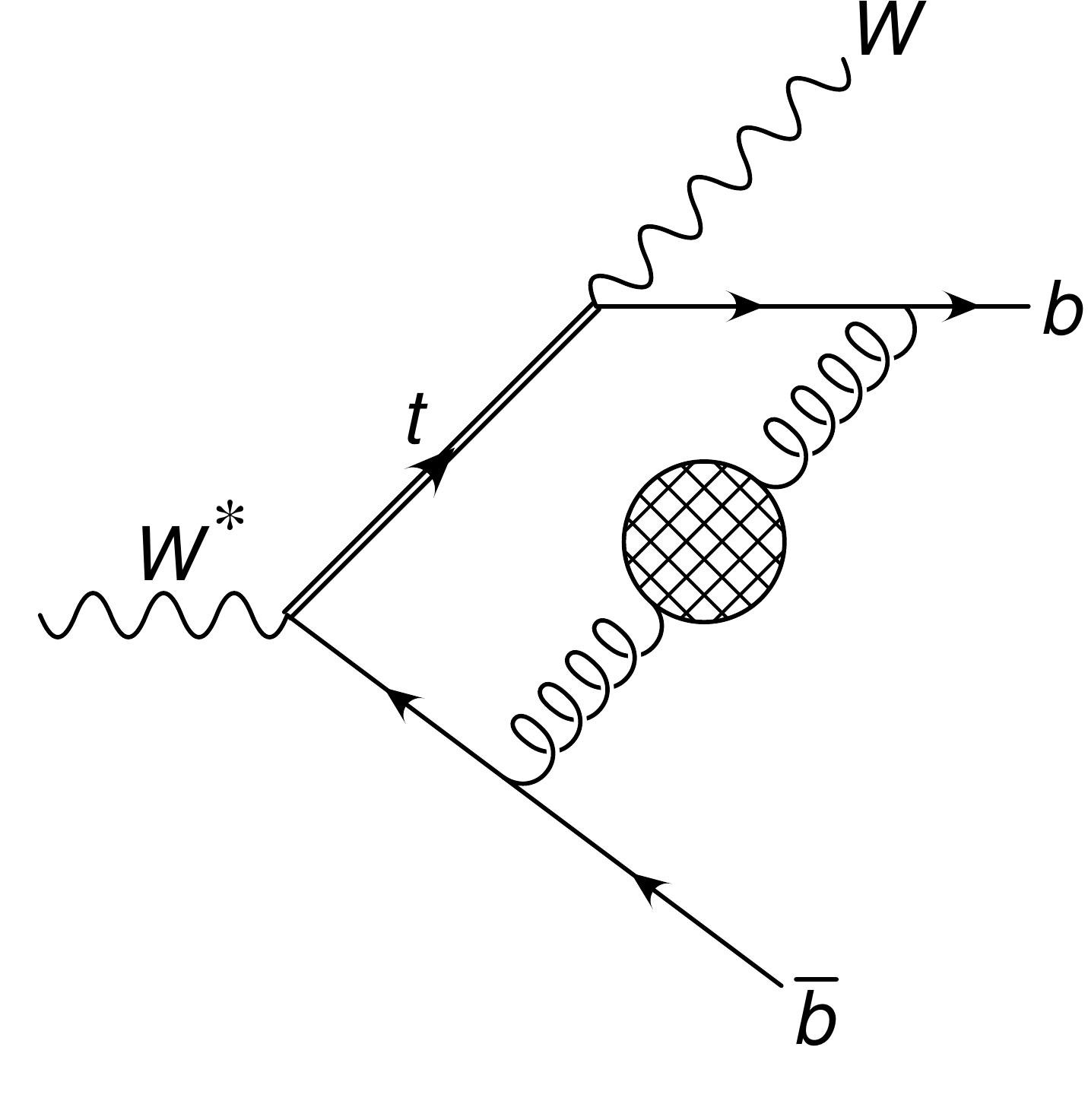}
 \includegraphics[width=0.24\textwidth]{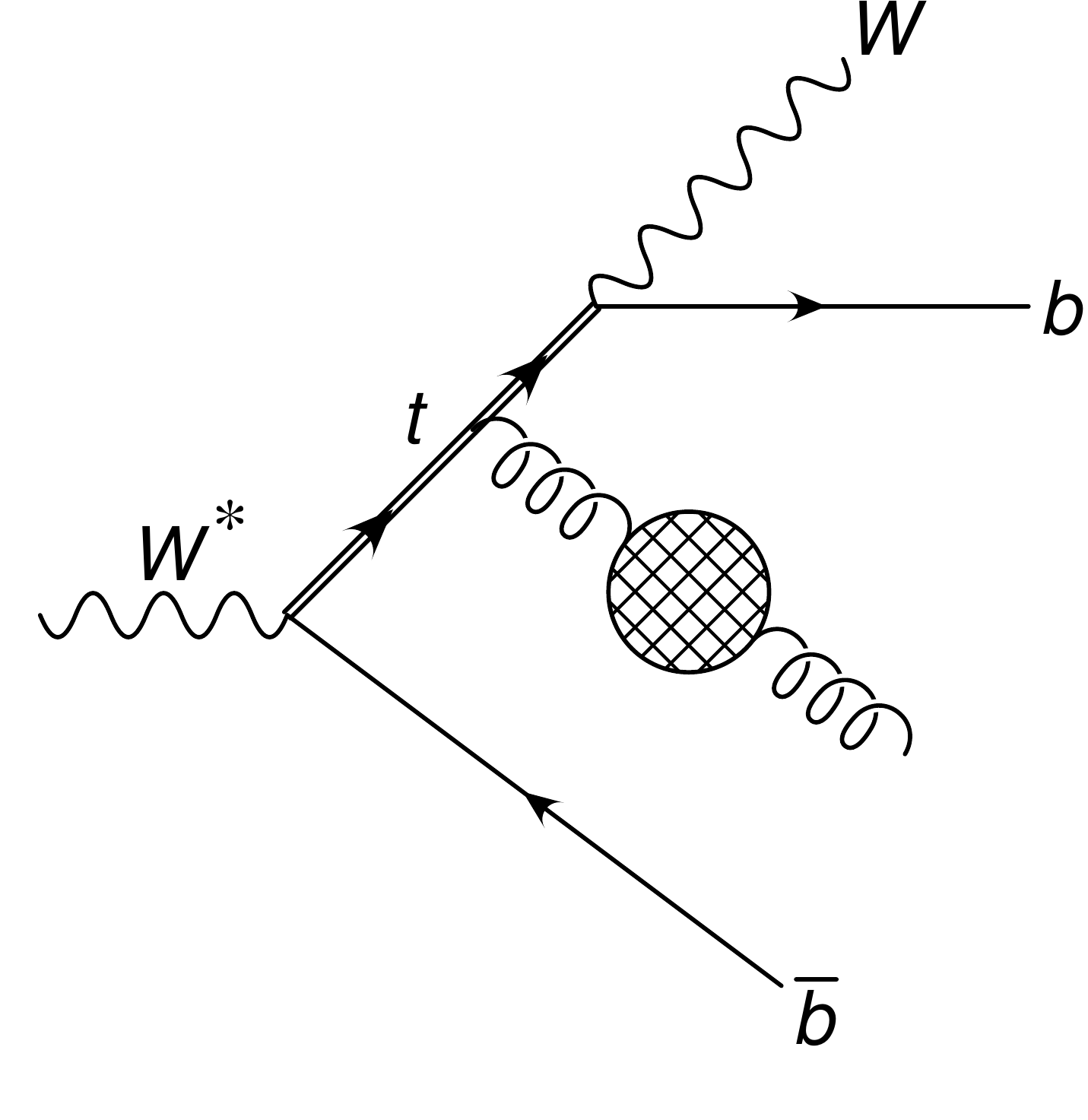}
 \includegraphics[width=0.24\textwidth]{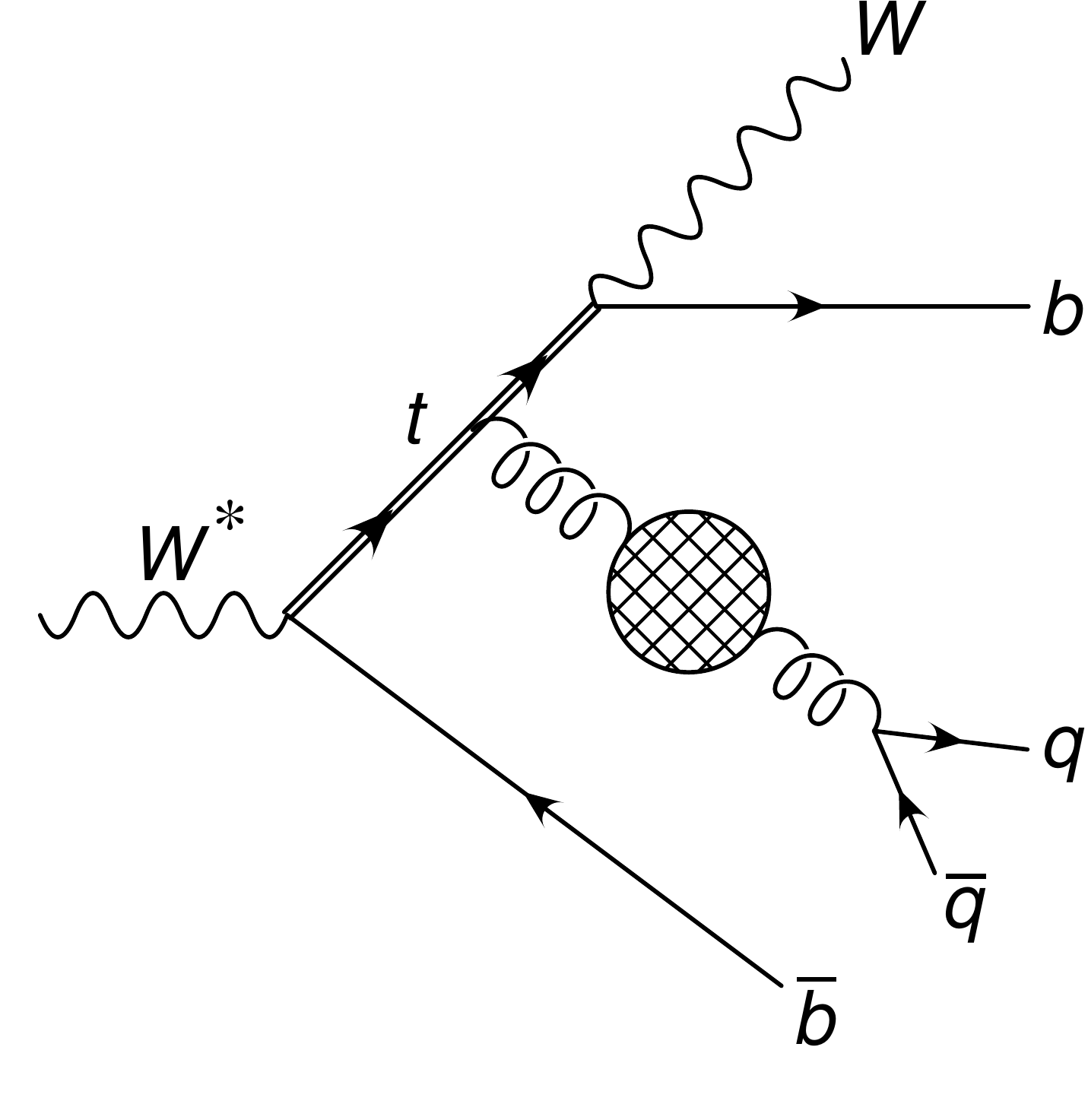}
 \includegraphics[width=0.6\textwidth]{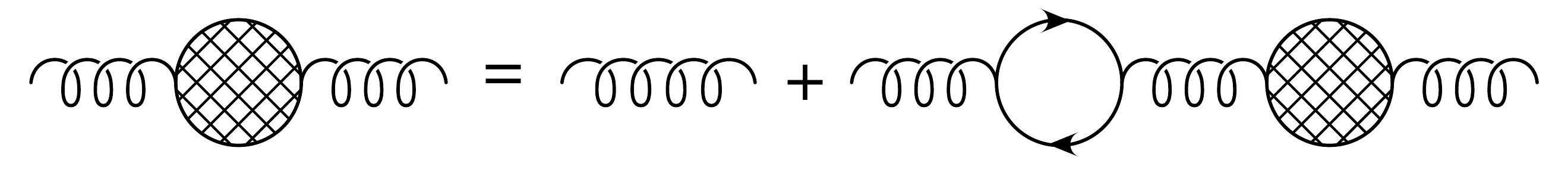}
\caption{Feynman diagram for the Born $W^*\to W b\bar{b}$ process,
 and samples of Feynman diagrams for the virtual contribution, for the
 real-emission contribution and for $W^*\to W b\bar{b}\, q\bar{q}$
 production.
 \label{fig:wbbbar}}
\end{figure}

In ref.~\cite{FerrarioRavasio:2018ubr} we have considered a simplified framework of top production and decay,
where a $t\bar{b}$ system is produced by the decay of a virtual $W$ with 300~GeV energy. The top
decays in turn into a $Wb$ system, with the $W$ on the mass shell. For simplicity,
we neglect the $b$ mass. We include
the strong corrections induced by the exchange or the emission of a gluon,
and all the corrections to the gluon propagator given by the insertion
of a light quark loop. The corresponding graphs are shown in fig.~\ref{fig:wbbbar},
that illustrate the set of diagrams that dominate in the
formal limit of a large number of flavours. In fact, they include
all strong corrections of order $\alpha_s(\alpha_s n_f)^n$ (where  $n_f$ is the number of light flavours)
for $n=0\ldots\infty$. The large $b_0$ approximation prescribes
that at the end of the calculation, in order to account for possible gluonic
corrections besides those due to fermion loops, one performs the replacement
\begin{equation}
  n_f \to n_f - \frac{11\,C_A}{4T_F},\quad\quad \mbox{with}\quad C_A=3,\quad T_F=\frac{1}{2}.
\end{equation}

The  large $b_0$ approximation gives a semi-quantitative estimate of the
leading large-order behaviour of the perturbative expansion due to
the so-called infra-red renormalons, i.e. factorial-growing coefficients
of the perturbative expansion, that for large orders behave as
$n! (2 b_0/k)^n \alpha_s^{n+1}$, where $k$ is a positive integer. The terms of the perturbative expansion
decrease as the order increases up to values of $n=\tilde{n}$ such that $\tilde{n}(2b_0/k) \alpha_s \approx 1$,
and for larger values they begin to increase. The size of the minimal terms is
\begin{equation}
  \tilde{n}! (2 \alpha_s b_0/k)^{\tilde{n}} \alpha_s^{\tilde{n}+1} \approx \alpha_s \sqrt{2\pi \tilde{n}} \exp[\tilde{n}(\log \tilde{n} -1)] \tilde{n}^{(-\tilde{n})}
  = \sqrt{\frac{k\pi\alpha_s}{b_0}} \exp\left[-\frac{k}{2b_0 \alpha_s}\right].
\end{equation}
Inserting the running coupling $\alpha_s=1/(b_0 \log \mu^2/\Lambda^2)$,
we see that the minimal term is of order $(\Lambda/\mu)^k$. Thus, renormalons are
both associated to the divergence of the perturbative expansion, and to power-suppressed
ambiguities in its resummation.

The case that interests us is $k=1$, since this can yield ambiguities in the top mass
measurements that are of order $\Lambda$, quite close to the present measurement errors,
while higher values of $k$ will lead to further suppressions by powers of $\Lambda/\mu$,
that, since $\mu\approx m$ ($m$ being the top mass) are fully negligible. In the following we will
use the term ``linear renormalon'' to denote $k=1$ renormalons.

Our results for any infrared-finite observable in our process has
the form
\begin{equation}\label{eq:renform}
  \langle O \rangle =  \langle O_b \rangle - \frac{1}{b_0\alpha_s}\int_0^\infty
  \frac{{\mathrm d}\lambda}{\pi} \frac{{\mathrm d} \widetilde{T}(\lambda)}{{\mathrm d}\lambda}
  \arctan\frac{\pi b_0 \alpha_s}{1+b_0\alpha_s \log \frac{\lambda^2}{\mu_C}},
\end{equation}
where $\langle O_b \rangle$ is the Born-level value of the observable and $\mu_C$
is proportional to the renormalization scale at which $\alpha_s$ is evaluated.
This formula has to be interpreted as a formal expansion in powers of
$\alpha_s$, so that we do not worry about the essential singularity arising when
the denominator of the argument of the arctangent vanishes.

Formulae of the
form of eq.~(\ref{eq:renform}) are known in the literature to arise in
the large-$b_0$ approximation~\cite{Beneke:1994qe}. The whole complexity of our calculation
is in the observable dependent function $\widetilde{T}(\lambda)$, that we can
compute with a semi-numerical method for any infrared-safe observable.
It turns out that in general $\widetilde{T}$, for small $\lambda$ goes to a constant,
plus a linear term in $\lambda$, plus terms of higher orders in $\lambda$, possibly
multiplied by logarithms of $\lambda$. It can be easily shown that the presence
of the linear term is associated with a linear renormalon.

Our calculation is performed in the pole mass scheme for the top mass. This guarantees
that the functions $\widetilde{T}(\lambda)$ vanishes fast enough for large $\lambda$
so that the integral in eq.~\ref{eq:renform} is convergent. It can be easily converted
into the corresponding expression for the $\bar{MS}$ scheme by using the
mass conversion formula evaluated in the large-$b_0$ limit~\cite{
Ball:1995ni}.
Since this formula is affected by linear renormalons,
the term linear in $\lambda$ in $\widetilde{T}$ is mass-scheme dependent, and in some
cases it vanishes in one of the two schemes.

We remark that the absence of a renormalon in a physical observable
when a short-distance mass scheme (like the $\overline{MS}$ scheme) is used, means
that no \emph{physical} renormalon is present in the observable. The same observable,
when expressed in terms of the pole mass, will have a renormalon that is
only due to the fact that it is expressed in terms of a quantity that has
a renormalon.

\section{The total cross section}
We begin by discussing the total cross section.
As also expected from general considerations,
we found no physical linear renormalon in this case, i.e. no factorial growth corresponding
to a linear renormalon when a short-distance mass scheme,
like the $\overline{\rm MS}$ one, is used.
This leads to a rather well-behaved perturbative expansion,
with the relative size of the terms of the expansion smaller than $10^{-5}$ already at the 4$^{\rm th}$
order, and with no visible minimum up to the 10$^{\rm th}$ order. On the other hand,
 in the pole mass scheme,
a linear renormalon appears, and
the minimal term of the expansion is reached at the $8^{\rm th}$ order, leading to
an ambiguity of relative order $5\times 10^{-4}$. This is of order of $0.1\,$GeV over the
mass of the top, as expected.

The benefit of using the $\overline{\rm MS}$ mass scheme for the total cross section
is greatly reduced if we need to impose acceptance cuts to identify our final state.
The same cross section,
with the restriction of requiring two separated $b$-jets with $R=0.1$ yields
a minimal term near $3\times 10^{-3}$ for both the pole mass and the $\overline{\rm MS}$ mass schemes,
while for the $r=0.5$ the minimal term is $-8\times 10^{-4}$ for the
pole scheme, and $-3.4\times 10^{-4}$ for the $\overline{\rm MS}$ scheme.
\section{The Reconstructed Top Mass}
We define the reconstructed top mass as the mass of the system comprising a
$b$ (not $\bar{b}$!) jet and an on-shell $W$. We find that this observable
has linear renormalons both
in the pole and in the $\overline{\rm MS}$ mass schemes,
with coefficients proportional to the inverse of the $R$ parameter used for
jets, as also found in other contexts~\cite{Korchemsky:1994is,Dasgupta:2007wa}.
In the  $\overline{\rm MS}$ scheme the perturbative
expansion begins with large positive corrections,
while that for the pole-mass scheme has large negative ones, that
are easily understood to arise from radiation
outside the jet cone from the $b$-jet. This radiation is also present in the
$\overline{\rm MS}$ scheme, that, however, has also a large positive correction,
due to the fact it grossly underestimates the pole position at the Born
level. This leads to a slightly smaller minimal terms for small to moderate
values of $R$. As $R$ becomes
larger, the out-of-cone radiation effect becomes less relevant,
and the large positive corrections to the pole position prevails,
leading to a larger minimal term with respect to the pole mass case.
It should be stressed that, in this case, the cancellation between renormalon effects
arising from two totally different contributions should not be taken as an indication
of a small overall ambiguity. Instead, since the cancellation is accidental,
one should consider the two contributions as independent sources of error.

As a last remark, we recall that, in the narrow width limit, one can in principle
separate the radiation arising in top decay from the one arising in production,
since they take place on very different time scales. If one was to define the reconstructed
top as the mass of the top decay products defined in this way, one would get
exactly the top pole mass. It is thus not surprising that, for relatively large $R$ values,
the renormalon ambiguity is reduced in the pole mass scheme. We can take this as an
indication that, for large $R$, we capture a large fraction of the top decay
products.

\section{Leptonic Observables}
As an example of leptonic observables we have taken $\langle E_W\rangle$,
i.e. the average energy of the $W$ boson.
This observable does not involve jets, and should thus be insensitive to renormalons
due to jet requirements. This observation has sometimes been used to advocate
leptonic observables with respect to hadronic ones, in spite of their smaller
sensitivity to the top mass.
Our results are summarized in table~\ref{tab:Wexp}.
\begin{table}[tb]
  \begin{center}
    {\small
    \begin{tabular}{c|c|c|c|c|}
  \cline{2-5}
  & \multicolumn{4}{|c|}{ $\displaystyle \phantom{\big|}\langle E_W \rangle\phantom{\big|}  $}
 \\ \cline{2-5}
  & \multicolumn{2}{|c|}{ \phantom{\Big|}pole scheme \phantom{\Big|}}& \multicolumn{2}{|c|}{$\overline{\rm MS}$ scheme}
 \\ \cline{1-5}
 \multicolumn{1}{|c|}{$\phantom{\Big|} i \phantom{\Big|}$}  & $c_i $ & $ c_i\,\alpha_s^i$  & $c_i $ & $ c_i \, \alpha_s^i$ 
 \\ \cline{1-5}
 \multicolumn{1}{|c|}{ $\phantom{\Big|}$ 0 $\phantom{\Big|}$}& $121.5818$ & $121.5818$& $120.8654$ & $120.8654$
 \\ \cline{1-5}
 \multicolumn{1}{|c|}{$\phantom{\Big|}$           1 $\phantom{\Big|}$}& $-1.435 \, (0) \times 10^{    1}$ & $-1.552 \, (0) \times 10^{    0}$ & $-7.192 \, (0) \times 10^{    0}$ & $-7.779 \, (0) \times 10^{   -1}$ 
 \\ \cline{1-5}
 \multicolumn{1}{|c|}{$\phantom{\Big|}$           2 $\phantom{\Big|}$}& $-4.97 \, (4) \times 10^{    1}$ & $-5.82 \, (4) \times 10^{   -1}$ & $-3.88 \, (4) \times 10^{    1}$ & $-4.54 \, (4) \times 10^{   -1}$ 
 \\ \cline{1-5}
 \multicolumn{1}{|c|}{$\phantom{\Big|}$           3 $\phantom{\Big|}$}& $-1.79 \, (5) \times 10^{    2}$ & $-2.26 \, (6) \times 10^{   -1}$ & $-1.45 \, (5) \times 10^{    2}$ & $-1.84 \, (6) \times 10^{   -1}$ 
 \\ \cline{1-5}
 \multicolumn{1}{|c|}{$\phantom{\Big|}$           4 $\phantom{\Big|}$}& $-6.9 \, (4) \times 10^{    2}$ & $-9.4 \, (6) \times 10^{   -2}$ & $-5.7 \, (4) \times 10^{    2}$ & $-7.8 \, (6) \times 10^{   -2}$ 
 \\ \cline{1-5}
 \multicolumn{1}{|c|}{$\phantom{\Big|}$           5 $\phantom{\Big|}$}& $-2.9 \, (3) \times 10^{    3}$ & $-4.4 \, (5) \times 10^{   -2}$ & $-2.4 \, (3) \times 10^{    3}$ & $-3.5 \, (5) \times 10^{   -2}$ 
 \\ \cline{1-5}
 \multicolumn{1}{|c|}{$\phantom{\Big|}$           6 $\phantom{\Big|}$}& $-1.4 \, (3) \times 10^{    4}$ & $-2.2 \, (4) \times 10^{   -2}$ & $-1.0 \, (3) \times 10^{    4}$ & $-1.7 \, (4) \times 10^{   -2}$ 
 \\ \cline{1-5}
 \multicolumn{1}{|c|}{$\phantom{\Big|}$           7 $\phantom{\Big|}$}& $-8 \, (2) \times 10^{    4}$ & $-1.3 \, (4) \times 10^{   -2}$ & $-5 \, (2) \times 10^{    4}$ & $-8 \, (4) \times 10^{   -3}$ 
 \\ \cline{1-5}
 \multicolumn{1}{|c|}{$\phantom{\Big|}$           8 $\phantom{\Big|}$}& $-5 \, (2) \times 10^{    5}$ & $-9 \, (4) \times 10^{   -3}$ & $-2 \, (2) \times 10^{    5}$ & $-4 \, (4) \times 10^{   -3}$ 
 \\ \cline{1-5}
 \multicolumn{1}{|c|}{$\phantom{\Big|}$           9 $\phantom{\Big|}$}& $-3 \, (2) \times 10^{    6}$ & $-7 \, (4) \times 10^{   -3}$ & $-1 \, (2) \times 10^{    6}$ & $-2 \, (4) \times 10^{   -3}$ 
 \\ \cline{1-5}
 \multicolumn{1}{|c|}{$\phantom{\Big|}$          10 $\phantom{\Big|}$}& $-3 \, (2) \times 10^{    7}$ & $-6 \, (5) \times 10^{   -3}$ & $0 \, (2) \times 10^{    6}$ & $-1 \, (5) \times 10^{   -4}$ 
 \\ \cline{1-5}
 \multicolumn{1}{|c|}{$\phantom{\Big|}$          11 $\phantom{\Big|}$}& $-3 \, (3) \times 10^{    8}$ & $-7 \, (6) \times 10^{   -3}$ & $0 \, (3) \times 10^{    6}$ & $0 \, (6) \times 10^{   -5}$ 
 \\ \cline{1-5}
 \multicolumn{1}{|c|}{$\phantom{\Big|}$          12 $\phantom{\Big|}$}& $-4 \, (3) \times 10^{    9}$ & $-9 \, (9) \times 10^{   -3}$ & $0 \, (3) \times 10^{    8}$ & $1 \, (9) \times 10^{   -3}$ 
 \\ \cline{1-5}
 \end{tabular}}
  \caption{Coefficients of the perturbative expansion
    of the average $W$-boson energy in the pole and $\overline{\rm MS}$-mass schemes.}
  \label{tab:Wexp}
  \end{center}
\end{table}

We draw the following conclusions:
  $\langle E_W\rangle$ has linear renormalons in both the  $\overline{\rm MS}$ and
  in the pole mass scheme, if the narrow-width limit is considered;
  in the finite width case, and in the $\overline{\rm MS}$ mass scheme,
  we have found evidence that the renormalon is screened
  at scales of the order of the top width. This means, in practice, that we observe
  the renormalon growth of the coefficients up to orders $n\approx \log(m/\Gamma)$,
  that in our case is near 5. For higher orders, the renormalon growth disappear.
\section{Conclusions}
In our study we have found that the largest sources of linear corrections
are those associated to jets, with a strength proportional to $1/R$. These
result is not unexpected. It is also likely that these
kind of corrections may be largely reduced if some procedure of jet calibration
is adopted.

A rather surprising result was found for the leptonic observables, where
physical linear renormalons are found in the narrow-width limit, and where evidence
for the screening of the physical renormalon due to the top finite width is found.
We were able also able to find a theoretical justification for both findings~\cite{FerrarioRavasio:2018ubr}.
We recall that, in $b$ decays, there are no linear renormalons associated with
leptonic observables~\cite{Bigi:1994em,Beneke:1994bc}.
This refers to leptonic observables are
computed in the bottom rest frame, and thus it is not in contrast with our findings.
If we compute
$\langle E_W\rangle$ in the top rest frame, we also find no linear renormalons.


\begin{thebibliography}{99}


\bibitem{CERN-LPCC-2018-03}
  R.Abdul Khalek \etal, CERN-LPCC-2018-03.
  
\bibitem{Beneke:1994qe}
  M.~Beneke and V.~M.~Braun,
  Phys.\ Lett.\ B {\bf 348} (1995) 513
  [hep-ph/9411229].

\bibitem{Ball:1995ni}
  P.~Ball, M.~Beneke and V.~M.~Braun,
  Nucl.\ Phys.\ B {\bf 452} (1995) 563
  [hep-ph/9502300].









\bibitem{FerrarioRavasio:2018ubr}
  S.~Ferrario Ravasio, P.~Nason and C.~Oleari,
  arXiv:1810.10931 [hep-ph].



  
\bibitem{Korchemsky:1994is}
  G.~P.~Korchemsky and G.~F.~Sterman,
  Nucl.\ Phys.\ B {\bf 437} (1995) 415
  [hep-ph/9411211].



\bibitem{Dasgupta:2007wa}
  M.~Dasgupta, L.~Magnea and G.~P.~Salam,
  JHEP {\bf 0802} (2008) 055
  [arXiv:0712.3014 [hep-ph]].




\bibitem{Bigi:1994em}
  I.~I.~Y.~Bigi, M.~A.~Shifman, N.~G.~Uraltsev and A.~I.~Vainshtein,
  Phys.\ Rev.\ D {\bf 50} (1994) 2234
  [hep-ph/9402360].

\bibitem{Beneke:1994bc}
  M.~Beneke, V.~M.~Braun and V.~I.~Zakharov,
  Phys.\ Rev.\ Lett.\  {\bf 73} (1994) 3058
  [hep-ph/9405304].



\end{thebibliography}
\end{document}